\documentclass{emulateapj}

\usepackage{amssymb}

\def\ts     {\thinspace}
\def\kms    {\ts km\ts s$^{-1}$}
\def\etal   {{\rm et\ts al.}}
\def\msol   {$M_{\odot}$}

\def\aco    {{\rm CO}($J$=1$\to$0)}

\def\dco    {{\rm CO}($J$=4$\to$3)}

\def\fco    {{\rm CO}($J$=6$\to$5)}

\def\smmja    {SMM\,J09431+4700}
\def\smmjb    {SMM\,J13120+4242}

\def\vla      {Very Large Array}
\def\evla     {Expanded Very Large Array}

\slugcomment{draft version \today, accepted for publication in the Astrophysical Journal Letters (EVLA Special Issue)}

\shorttitle{\aco\ Emission in $z$$\simeq$3.4 Submillimeter Galaxies}
\shortauthors{Riechers et al.}

\begin{document}

\title{
Extended Cold Molecular Gas Reservoirs in $z$$\simeq$3.4 Submillimeter Galaxies
}

\author{Dominik A.\ Riechers\altaffilmark{1}, Jacqueline Hodge\altaffilmark{2},
Fabian Walter\altaffilmark{2}, Christopher L.\ Carilli\altaffilmark{3}, 
and Frank Bertoldi\altaffilmark{4}}

\altaffiltext{1}{Astronomy Department, California Institute of
  Technology, MC 249-17, 1200 East California Boulevard, Pasadena, CA
  91125, USA; dr@caltech.edu}

\altaffiltext{2}{Max-Planck-Institut f\"ur Astronomie, K\"onigstuhl 17, D-69117 Heidelberg, Germany}

\altaffiltext{3}{National Radio Astronomy Observatory, PO Box O, Socorro, NM 87801, USA}

\altaffiltext{4}{Argelander-Institut f\"ur Astronomie, Universit\"at
  Bonn, Auf dem H\"ugel 71, D-53121 Bonn, Germany}


\begin{abstract}

We report the detection of spatially resolved \aco\ emission in the
$z$$\sim$3.4 submillimeter galaxies (SMGs) SMM\,J09431+4700 and
SMM\,J13120+4242, using the Expanded Very Large Array
(EVLA). SMM\,J09431+4700 is resolved into the two previously reported
millimeter sources H6 and H7, separated by $\sim$30\,kpc in
projection.  We derive \aco\ line luminosities of $L'_{\rm CO(1-0)}\ =\
(2.49 \pm 0.86)\,{\rm and}\,(5.82 \pm 1.22) \times
10^{10}\,$K\,\kms\,pc$^2$ for H6 and H7, and $L'_{\rm CO(1-0)}\ =\ (23.4
\pm 4.1) \times 10^{10}\,$K\,\kms\,pc$^2$ for SMM\,J13120+4242. 
These are $\sim$1.5--4.5$\times$ higher than what is expected from
simple excitation modeling of higher-$J$ CO lines, suggesting the
presence of copious amounts of low-excitation gas. This is supported
by the finding that the \aco\ line in SMM\,J13120+4242, the system
with lowest CO excitation, appears to have a broader profile and more
extended spatial structure than seen in higher-$J$ CO lines (which is
less prominently seen in SMM\,J09431+4700).  Based on $L'_{\rm
CO(1-0)}$ and excitation modeling, we find $M_{\rm gas}$=2.0--4.3 and
4.7--12.7$\times$10$^{10}$\,\msol\ for H6 and H7, and $M_{\rm
gas}$=18.7--69.4$\times$10$^{10}$\,\msol\ for SMM\,J13120+4242.  The
observed \aco\ properties are consistent with the picture that
SMM\,J09431+4700 represents an early-stage, gas-rich major merger, and
that SMM\,J13120+4242 represents such a system in an advanced stage.
This study thus highlights the importance of spatially and dynamically
resolved \aco\ observations of SMGs to further understand the gas
physics that drive star formation in these distant galaxies, which
becomes possible only now that the EVLA rises to its full
capabilities.

\end{abstract}


\section{Introduction}

With star formation rates of typically $>$500\,\msol\,yr$^{-1}$,
submillimeter galaxies (SMGs) are the most intense starbursts known at
early cosmic times (see review by Blain et al.\ \citeyear{bla02}). A
significant fraction of the stellar mass assembly in massive galaxies
at high redshift is thought to occur at early epochs ($z$$>$2) on
rapid timescales ($<$100\,Myr) through a dusty, gas-rich SMG phase
(e.g., Greve et al.\ \citeyear{gre05}), making SMGs the likely
progenitors of massive galaxies in the present day universe.

A particularly useful probe to better understand the physical
conditions for star formation in these extreme starbursts are studies
of the molecular interstellar medium that fuels the formation of
stars. Dense molecular gas traced through CO was detected in 34 SMGs
to date, typically revealing gas reservoirs of few 10$^{10}$\,\msol\
(see review by Solomon \& Vanden Bout \citeyear{sv05}). However, most
of these detections are of CO rotational lines from $J$$\geq$3
transitions, which may bias these studies toward highly-excited gas,
and do not necessarily trace the entire molecular gas reservoir as
traced by \aco. Indeed, recent studies hint at the presence of
substantial amounts of low-excitation gas in some SMGs (e.g., Hainline
et al.\ \citeyear{hai06}; Carilli et al.\ \citeyear{car10}; Harris et
al.\ \citeyear{har10}; Ivison et al.\ \citeyear{ivi10}; Riechers et
al.\ \citeyear{rie10}). Unfortunately, previous \aco\ studies in SMGs
(which do not suffer from excitation bias) were either carried out
with single-dish telesopes lacking spatial information, or with the
old correlator of the \vla\ (VLA) lacking detailed spectral
information.

To overcome the limitations of previous studies, we have initiated a
systematic study of the \aco\ content of SMGs and other galaxy
populations with the \evla\ (EVLA; Perley et al.\
\citeyear{per11}). We here report early results on two $z$$\sim$3.4
SMGs that appear to trace different stages of gas-rich, major mergers
only $\sim$1.9\,Gyr after the Big Bang. \smmja\ ($z$=3.346) hosts two
SMGs detected in $J$$\geq$4 CO emission, separated by $\sim$30\,kpc in
projection, likely representing an early phase of a merger (e.g.,
Tacconi et al.\ \citeyear{tac06}; Engel et al.\ \citeyear{eng10}). 
\smmjb\ ($z$=3.408) shows a complex, disturbed structure in \fco, 
suggesting that it may represent an advanced-stage merger (Engel et
al.\ \citeyear{eng10}).  We use a concordance, flat $\Lambda$CDM
cosmology throughout, with $H_0$=71\,\kms\,Mpc$^{-1}$, $\Omega_{\rm
M}$=0.27, and $\Omega_{\Lambda}$=0.73 (Spergel \etal\
\citeyear{spe03}, \citeyear{spe07}).

\begin{figure*}
\vspace{-6mm}

\epsscale{1.0}
\plotone{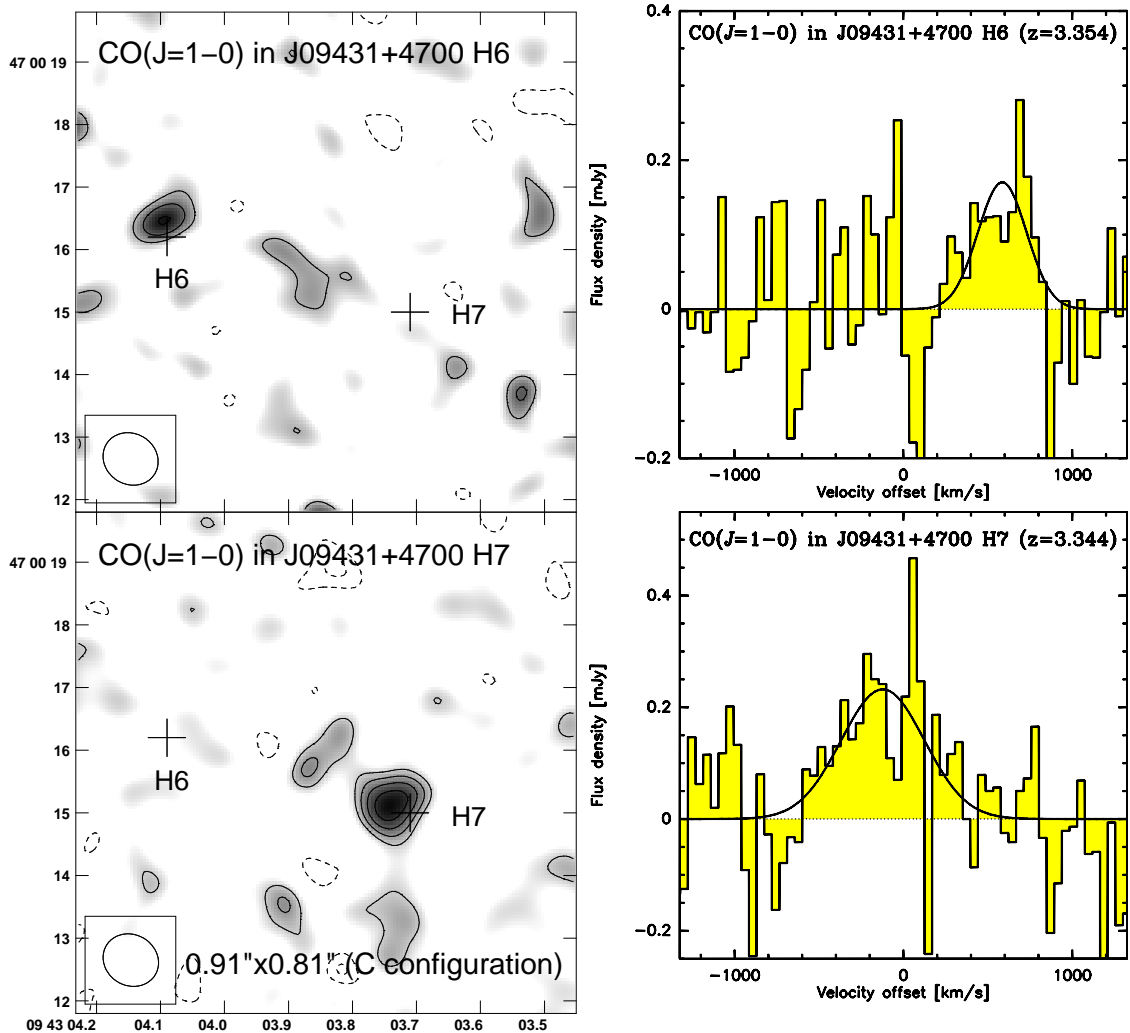}
\vspace{-12mm}

\caption{EVLA maps (left) and spectra (right) of \aco\ emission toward the $z$=3.346 SMG J09431+4700 H6 (top) and H7 (bottom). The resolution of both maps is 0.91$''$$\times$0.81$''$. The crosses indicate continuum positions reported by Tacconi et al.\ (\citeyear{tac06}). Contours are shown in steps of 1$\sigma$=30\,$\mu$Jy\,beam$^{-1}$, starting at $\pm$2$\sigma$. The spectra (histograms) are shown at 45\,\kms\ (4\,MHz) resolution. Zero velocity corresponds to $z$=3.346. The curves indicate Gaussian fits to the spectra. Top:\ Emission toward H6, averaged over 475\,\kms\ (42\,MHz; central velocity:\ 585\,\kms ). Bottom:\ Emission toward H7, averaged over 543\,\kms\ (48\,MHz; central velocity:\ $-$123\,\kms ).\label{f2}}
%
\end{figure*}

\section{Observations}

We observed the \aco\ ($\nu_{\rm rest} = 115.2712$\,GHz) emission line
toward \smmja\ and \smmjb, using the EVLA.  At $z$=3.346 and 3.408,
this line is redshifted to 26.5235 and 26.1505\,GHz ($\sim$1.14\,cm;
$K$ band).  Observations were carried out under good weather
conditions in D array on 2010 May 19 and in C array between 2010
December 16 and 2011 January 19, resulting in 4.0\,hr (2.0\,hr) total
(on-source) observing time for \smmja\ (C array), and 7.5\,hr
(4.6\,hr) for \smmjb\ (C+D array).  The nearby quasars J0920+4441 and
J1312+4828 were observed every 4 to 7\,minutes for pointing, secondary
amplitude and phase calibration. For primary flux calibration, the
standard calibrator 3C286 was observed, leading to a calibration that
is accurate within $\sim$10\%.  Observations were set up using a total
bandwidth of 252\,MHz (dual polarization; corresponding to
$\sim$2900\,\kms\ at 1.14\,cm) with the WIDAR correlator.

For data reduction and analysis, the AIPS package was used.  All data
were mapped using `natural' weighting.  The data result in a final rms
of 105 and 70\,$\mu$Jy\,beam$^{-1}$ per $\sim$45\,\kms\ (4\,MHz)
channel for \smmja\ and \smmjb, respectively.  Maps
of the velocity-integrated CO $J$=1$\to$0 line emission yield
synthesized clean beam sizes of 0.91$''$$\times$0.81$''$ and
1.01$''$$\times$0.72$''$ at rms noise levels of 30 and
12\,$\mu$Jy\,beam$^{-1}$ over 543 and 1513\,\kms\ (48 and 132\,MHz)
for \smmja\ and \smmjb.

\begin{figure*}
\vspace{-6mm}

\epsscale{1.0}
\plotone{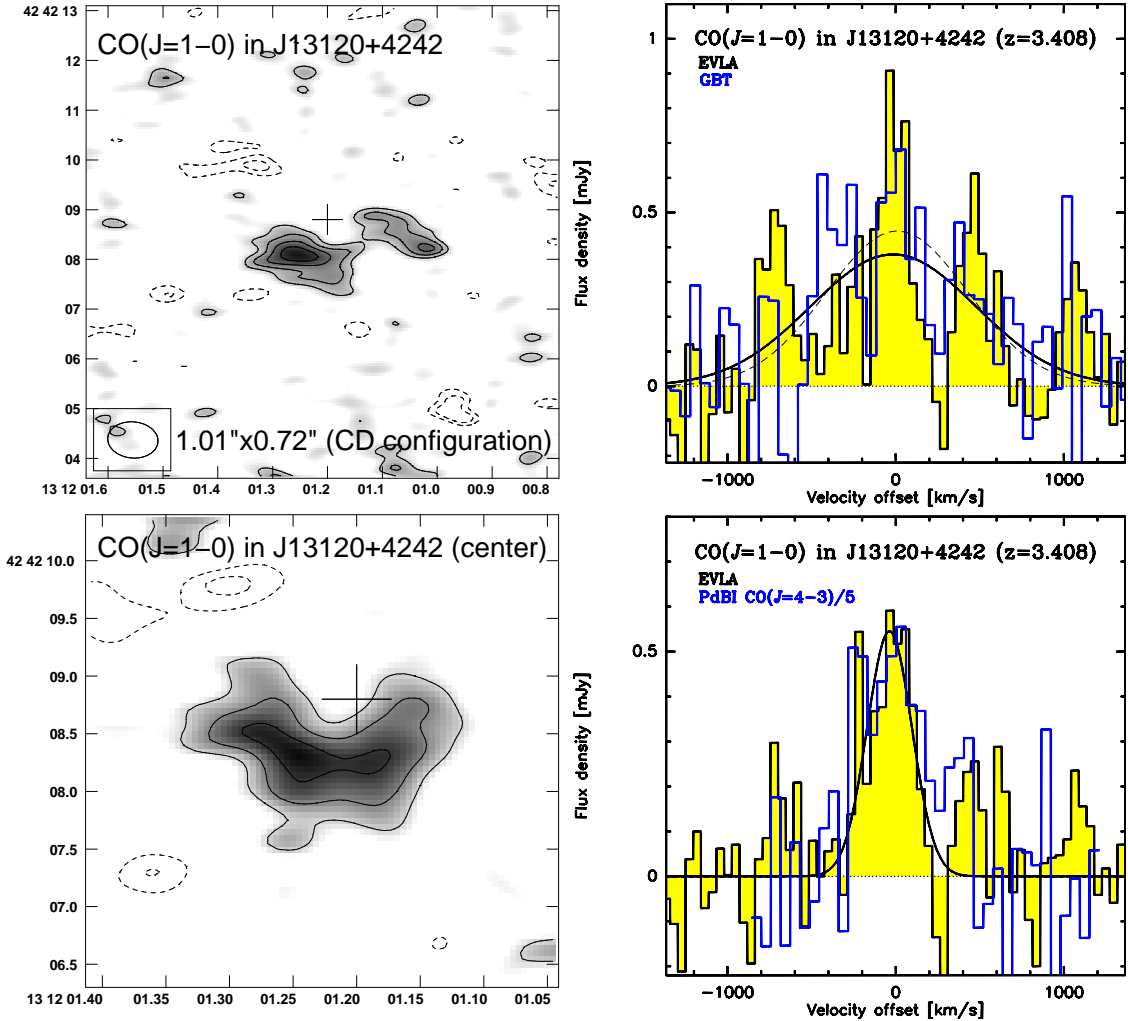}
\vspace{-10mm}

\caption{EVLA maps (left) and spectra (right) of \aco\ emission toward the $z$=3.408 SMG J13120+4242. The resolution of both maps is 1.01$''$$\times$0.72$''$. The cross indicates the center of the \dco\ emission detected at $\sim$6$''$ resolution (Greve et al.\ \citeyear{gre05}). Contours are shown in steps of 1$\sigma$=12 and 29\,$\mu$Jy\,beam$^{-1}$ (top/bottom), starting at $\pm$2$\sigma$. The spectra (histograms) are shown at 46\,\kms\ (4\,MHz) resolution, and are extracted over the emission regions delimited by the 2$\sigma$ contours in the maps to the left. Zero velocity corresponds to $z$=3.408. The solid curves indicate Gaussian fits to the spectra. Top:\ Total emission over 1513\,\kms\ (132\,MHz). The empty histogram and dashed curve in the right panel indicate the single-dish measurement with the Green Bank Telescope at $\sim$28$''$ resolution (Hainline et al.\ \citeyear{hai06}). Bottom:\ Emission over the central 344\,\kms\ (30\,MHz). The empty histogram shows the \dco\ emission line, scaled down by a factor of 5 in flux density (Greve et al.\ \citeyear{gre05}). \label{f1}}
%
\end{figure*}

\section{Results}

\begin{figure*}
\vspace{-2mm}
\epsscale{1.15}
\plotone{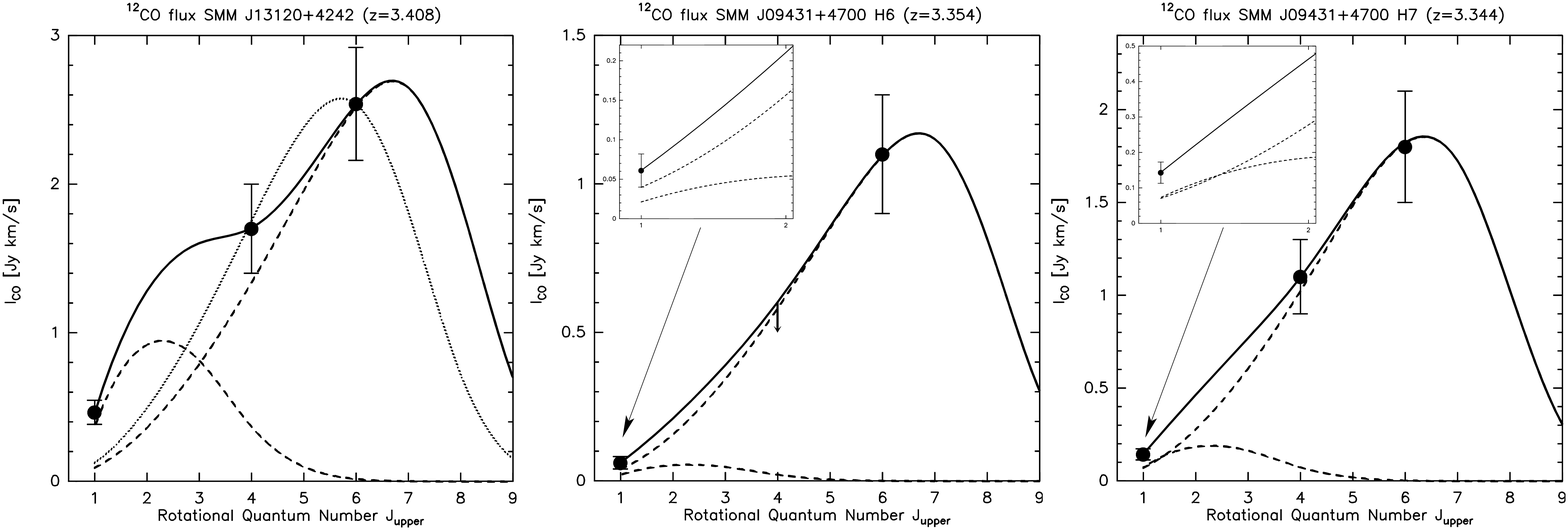}

\caption{CO excitation ladders (points) and LVG models (lines) for SMM\,J13120+4242 and SMM\,09431+4700 H6 and H7. The \dco\ and \fco\ data are adopted from Greve et al.\ (\citeyear{gre05}), Tacconi et al.\ (\citeyear{tac06}) and Engel et al.\ (\citeyear{eng10}). The insets show zoomed-in versions of the model fits close to the \aco\ line. The data for all sources are fit well by two-component LVG models, as indicated by the solid lines.
The dashed lines indicate the two components of each model,
represented by a low-excitation component with a kinetic temperature
of $T_{\rm kin}$=25\,K and a gas density of $\rho_{\rm
gas}$=10$^{2.5}$\,cm$^{-3}$, and a high excitation component with
$T_{\rm kin}$=40\,K and a gas density of $\rho_{\rm
gas}$=10$^{4.3}$\,cm$^{-3}$ (this component has a 1.26$\times$ lower
$\rho_{\rm gas}$ for H7). The high-excitation components contribute
20\%, 65\%, and 49\% to the observed \aco\ intensities. For
comparison, the dotted line in the left panel shows a single-component
fit to the $J$$>$1 lines with $T_{\rm kin}$=40\,K and $\rho_{\rm
gas}$=10$^{4.0}$\,cm$^{-3}$. \label{f3}}
%
\end{figure*}

\subsection{SMM\,J09431+4700}

We detected spatially resolved \aco\ emission toward
\smmja. The emission is resolved into two components that
are separated by $\sim$30\,kpc in projection, and are identified with
the radio/millimeter continuum sources H6 and H7 (Fig.~\ref{f2}).
Component H7 appears marginally resolved at the $\sim$6.5\,kpc
resolution of our observations, suggesting an approximate source
radius of $\sim$3$\pm$1\,kpc. This suggests that the \aco\ line
emission may be somewhat more extended than what is observed in
higher-$J$ lines (1.85$\pm$0.60 and 1.13$\pm$0.23\,kpc in the CO
$J$=4$\to$3 and 6$\to$5 lines; Engel et al.\ \citeyear{eng10}).
Component H6 appears unresolved at the moderate achieved
signal--to--noise ratio.  There is marginal evidence for some low
surface brightness emission in between H6 and H7, but more sensitive
observations are required to investigate this in more detail. We do
not detect continuum emission toward either component down to a
3$\sigma$ limit of 44\,$\mu$Jy\,beam$^{-1}$ at 1.1\,cm (rest-frame
2.6\,mm), consistent with the spectral energy distribution of this
source.

From Gaussian fitting to the line profiles of H6 and H7, we obtain
line peak strengths of $S_{\nu}$=170$\pm$58 and 232$\pm$47\,$\mu$Jy at
line FWHMs of d$v$=339$\pm$136 and 580$\pm$143\,\kms, centered at
redshifts of $z$=3.3545$\pm$0.0009 and $z$=3.3442$\pm$0.0009,
respectively. The line widths correspond to 1.9$\times$ (H6) and
1.5$\times$ (H7) those measured in \fco\ (Engel et al.\
\citeyear{eng10}). The line parameters for H6 and H7 correspond to 
velocity-integrated emission line strengths of $I_{\rm
CO(1-0)}$=0.061$\pm$0.021 and 0.143$\pm$0.030\,Jy\,\kms, suggesting
\aco\ line luminosities of $L'_{\rm CO}$=(2.49$\pm$0.86) and 
(5.82$\pm$1.22) $\times$ 10$^{10}$\,($\mu_{\rm
L}$/1.2)$^{-1}$\,K\,\kms\,pc$^{2}$ (where $\mu_{\rm L}$=1.2 is the
lensing magnification factor; this SMG is located behind the galaxy
cluster A851; Cowie et al.\ \citeyear{cow02}).

\subsection{SMM\,J13120+4242}

We detected spatially resolved \aco\ emission toward
\smmjb. The emission peaks on a
(2.0$''$$\pm$0.6$''$)$\times$(0.9$''$$\pm$0.4$''$)
(15.0$\times$6.7\,kpc$^2$) size region (Fig.~\ref{f1}, bottom). The
line profile averaged over this region is consistent with that seen in
the \dco\ line (empty histogram in Fig.~\ref{f1}, bottom; Greve et
al.\ \citeyear{gre05}), and its complex morphological structure is
consistent with what is seen in \fco\ line emission (Engel et al.\
\citeyear{eng10}).  This component, however, carries only
$\lesssim$40\% of the observed \aco\ line flux, and appears to be
embedded in a kinematically a few times broader component that extends
out to $\gtrsim$2$\times$ larger scales (Fig.~\ref{f1}, top). The line
profile over this whole extended structure appears to consist of
multiple peaks. For consistency, we fit the line with a Gaussian
profile, yielding $S_{\nu}$=380$\pm$65\,$\mu$Jy and
d$v$=1153$\pm$235\,\kms, and $I_{\rm
CO(1-0)}$=0.464$\pm$0.081\,Jy\,\kms, centered at
$z$=3.4078$\pm$0.0014. Within the relative uncertainties, the line
profile and parameters are consistent with a single-dish measurement
of this line, carried out with the Green Bank Telescope at 28$''$
resolution (empty histogram and dashed curve in Fig.~\ref{f1}, top;
Hainline et al.\ \citeyear{hai06}). This suggests that the large
apparent spatial extent of the \aco\ emission is real, and that
virtually all line flux is recovered in the EVLA map. The line FWHM
corresponds to 2.2$\times$ and 1.3$\times$ those measured in the \dco\
and \fco\ lines. We do not detect any rest-frame 2.6\,mm continuum
emission down to a 3$\sigma$ limit of 60\,$\mu$Jy\,beam$^{-1}$,
consistent with the spectral energy distribution of this
source. Averaged over the full region where \aco\ line emission is
detected, this suggests a conservative 3$\sigma$ limit of
$\sim$270\,$\mu$Jy. From the line intensity, we derive $L'_{\rm
CO}$=(23.4$\pm$4.1) $\times$ 10$^{10}$\,K\,\kms\,pc$^{2}$.

\begin{figure*}
\vspace{-17.5mm}

\epsscale{1.23}
\plotone{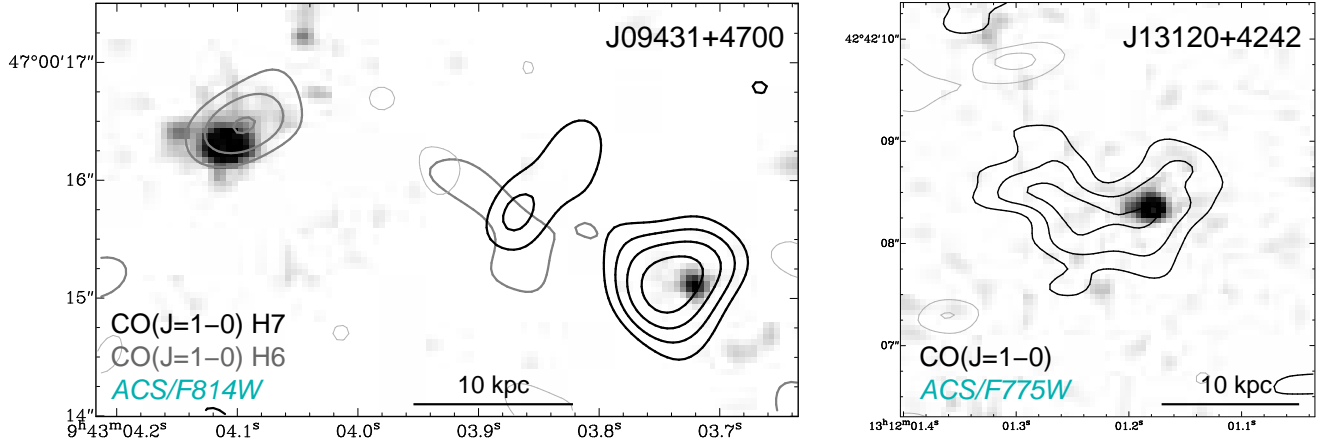}
\vspace{-17.5mm}

\caption{Overlays of the \aco\ emission (contours, as in Figures \ref{f2} and \ref{f1}) toward J09431+4700 (left) and the center of J13120+4242 (right) on {\em Hubble Space Telescope} Advanced Camera for Surveys (ACS) F814W (McGrath et al.\ \citeyear{mcg08}) and F775W images (gray scale; rest-frame $\sim$180\,nm; smoothed with a 0.1$''$ Gaussian kernel; Hainline et al.\ \citeyear{hai06}). The gray and black contours in the left panel indicate the velocity ranges for H6 and H7 as in Fig.~\ref{f2}.
 \label{f4}}
%
\end{figure*}

\section{Analysis}

\subsection{Line Excitation Modeling}

Besides the EVLA \aco\ observations presented here, \smmja\
and \smmjb\ were studied in \dco\ and \fco\ line emission
(see Engel et al.\ \citeyear{eng10} for a summary of previous
studies).  Based on the observed CO excitation ladders, we can
constrain the line radiative transfer through Large Velocity Gradient
(LVG) models, treating the gas kinetic temperature and density as free
parameters.  For all calculations, the H$_2$ ortho--to--para ratio was
fixed to 3:1, the cosmic microwave background temperature was fixed to
11.84 and 12.01\,K (at $z$=3.346 and 3.408), and the Flower
(\citeyear{flo01}) CO collision rates were used. We adopted a CO
abundance per velocity gradient of [CO]/(${\rm d}v/{\rm d}r) = 1
\times 10^{-5}\,{\rm pc}\,$(\kms)$^{-1}$ (e.g., Wei\ss\ \etal\
\citeyear{wei05b}, \citeyear{wei07}; Riechers \etal\ \citeyear{rie06}). 
Components H6 and H7 of \smmja\ are fitted individually.

The sources are poorly fit by single-component models, which
underpredict the observed \aco\ fluxes by factors of $\sim$1.5--4.5
when fitted to the higher-$J$ lines.\footnote{Given the limited
constraints on H6, this source can be marginally fit with a single
component.} The data for all sources can be fitted reasonably well
with two gas components, which are represented by a `diffuse',
low-excitation component with a kinetic temperature of $T_{\rm
kin}$=25\,K and a gas density of $\rho_{\rm
gas}$=10$^{2.5}$\,cm$^{-3}$, and a more `dense', high-excitation
component with $T_{\rm kin}$=40\,K (comparable to the dust temperature
in \smmjb; Kov\'acs et al.\
\citeyear{kov06}) and $\rho_{\rm gas}$=10$^{4.3}$\,cm$^{-3}$ in our
models ($\rho_{\rm gas}$=10$^{4.2}$\,cm$^{-3}$ is used for H7, as it
provides a slightly better fit). The two gas components differ in
relative strength between the sources (Fig.~\ref{f3}). For H6 and H7,
we find that the dense gas components have surface filling factors of
13\% and 7\% relative to the low-excitation component. For
\smmjb, we find that only $\sim$2\% of the surface area
associated with emission from the low-excitation component also shows
emission from the dense gas component. This suggests that only a small
fraction of the volume in these galaxies occupied by gas is associated
with dense regions. Also, these model parameters suggest that the
low-excitation gas components fill a dominant fraction of the area
over which the sources are resolved in these observations.

The physical properties of the low-excitation gas components in these
models are comparable to those of the gas in nearby spiral galaxies
and `normal' high-$z$ star-forming galaxies (e.g., Dannerbauer \etal\
\citeyear{dan09}) and the low-excitation components found in 
other SMGs (e.g., Carilli et al.\ \citeyear{car10}; Riechers et al.\
\citeyear{rie10}). The properties of the high-excitation components 
are also comparable to what is found in other SMGs, ultra-luminous
infrared galaxy (ULIRG) nuclei, and high-$z$ FIR-luminous quasars
(e.g., Riechers et al.\ \citeyear{rie06}, \citeyear{rie09}; Wei\ss\ et
al.\ \citeyear{wei05}). However, it appears that the relative
fractions of low- and high-excitation gas differ between individual
systems, perhaps due to different evolutionary stages.

\subsection{Gas Masses and Surface Densities, Dynamical Masses, Gas Mass Fractions}

Based on the observed \aco\ line luminosities, we can derive the total
masses of the gas reservoirs in our targets. For SMGs, a ULIRG-like
conversion factor of $\alpha_{\rm CO}$=$M_{\rm gas}$/$L'_{\rm
CO}$=0.8\,$M_\odot$\,(K\,\kms\,pc$^2$)$^{-1}$ is commonly adopted
(e.g., Downes \& Solomon \citeyear{ds98}; Tacconi et al.\
\citeyear{tac08}). This suggests $M_{\rm gas}$=2.0 and
4.7$\times$10$^{10}$\,($\mu_{\rm L}$/1.2)$^{-1}$\,$M_\odot$ for
\smmja\ H6 and H7, and $M_{\rm gas}$=18.7$\times$10$^{10}$\,$M_\odot$
for \smmjb, respectively. However, a low, ULIRG-like conversion factor
may not be appropriate for the low-excitation gas components suggested
by our models. Assuming $\alpha_{\rm
CO}$=3.5\,$M_\odot$\,(K\,\kms\,pc$^2$)$^{-1}$ for these components (as
suggested for gas components with comparable excitation found in
disk-like high-$z$ galaxies; e.g., Daddi et al.\ \citeyear{dad10}), we
find $M_{\rm gas}$=4.3 and 12.7$\times$10$^{10}$\,($\mu_{\rm
L}$/1.2)$^{-1}$\,$M_\odot$ for \smmja\ H6 and H7, and $M_{\rm
gas}$=69.4$\times$10$^{10}$\,$M_\odot$ for \smmjb, respectively.

Assuming FWHM radii of 3\,kpc for \smmja\ H6 and H7, the gas
masses correspond to average gas surface densities of $\Sigma_{\rm
gas}$=700--1500 and 1600--4500\,$M_\odot$\,pc$^{-2}$. The \aco\
emission in \smmjb\ is distributed over a projected area of
$\sim$200\,kpc$^2$, corresponding to $\Sigma_{\rm
gas}$=960--3600\,$M_\odot$\,pc$^{-2}$.

\smmjb\ has a stellar mass of
$M_\star$=10.2--13.5$\times$10$^{10}$\,$M_\odot$ (Engel et al.\
\citeyear{eng10}). This yields a baryonic gas mass fraction of 
$f_{\rm gas}^{\rm bary}$=$M_{\rm gas}$/($M_{\rm
gas}$+$M_{\star}$)=0.58--0.87, where the spread of values represents
the difference in conversion factors as outlined above.

Using the ``isotropic virial estimator'' (e.g., Tacconi et al.\
\citeyear{tac08}; Engel et al.\ \citeyear{eng10}), we can also 
estimate the dynamical masses of our targets. Assuming FWHM radii of
3\,kpc for \smmja\ H6 and H7, we find $M_{\rm dyn}$=9.7 and
28.3$\times$10$^{10}$\,$M_\odot$, respectively. This corresponds to
gas mass fractions $f_{\rm gas}$=$M_{\rm gas}$/$M_{\rm
dyn}$=0.21--0.45 and 0.16--0.45 for H6 and H7, where the spread of
values represents the use of different conversion factors as outlined
above. This lies within the typical range of values estimated for SMGs
(e.g., Tacconi et al.\ \citeyear{tac06}, \citeyear{tac08}). The \aco\
emission in \smmjb\ is spread over an area with a characteristic
radius of 7.9\,kpc.  Assuming that the multiple peaks can be described
by a single potential that is represented correctly by the FWHM on the
Gaussian fit used above, we find $M_{\rm
dyn}$=290$\times$10$^{10}$\,$M_\odot$ for \smmjb. This corresponds to
$f_{\rm gas}$=0.06--0.24. Under the assumption that the system is
gravitationally bound, and that $M_{\rm gas}$+$M_{\star}$ corresponds
to the bulk of baryonic matter in this system (the contributions of
dust and black hole mass are expected to be minor), this would imply
that this SMG is dominated by dark matter.  However, given the complex
velocity structure of this system, we consider it likely that $M_{\rm
dyn}$ is overestimated by a factor of 2--3 with this simplified
approach. Thus, we consider the gas fraction for \smmjb\ a lower
limit.

\subsection{Origin of the Molecular Line Emission}

\smmjb\ and both components of \smmja\ are
associated with galaxies detected with the {\em Hubble Space
Telescope} in rest-frame optical/ultraviolet light
(Fig.~\ref{f4}). H7, the more CO-luminous component of
\smmjb, is not significantly detected in $R$-band images of this source
(Tacconi et al.\ \citeyear{tac06}), which may indicate that it is more
obscured than H6. The CO emission centroids in H6 and H7 are offset by
$\sim$1.5\,kpc (0.2$''$) from the centers of the optical emission, but
are clearly associated with fainter, potentially more obscured regions
in the galaxies. The optical emission in \smmjb\ appears to be
associated with one of the CO peaks in the central region, and thus,
potentially with one component of a major merger (or a region of low
dust obscuration).

\section{Discussion}

We have detected spatially resolved \aco\ emission toward the
$z$$\sim$3.4 SMGs \smmja\ and \smmjb. We resolve
the gas reservoirs down to $\lesssim$5\,kpc scales (given the sizes of
our synthesized beams). \aco\ emission in \smmja\ is detected
toward both radio/millimeter continuum sources H6 and H7, which have
been detected in higher-$J$ CO lines in earlier work. There is
marginal evidence for additional \aco\ emission in between the two
sources, which are separated by $\sim$30\,kpc in projection. This, and
their proximity in redshift (d$z$$\lesssim$0.01) is consistent with
the picture that this SMG is undergoing an early stage massive,
gas-rich merger (see also Engel et al.\
\citeyear{eng10}). The \aco\ line in H7, the more CO-luminous
component, appears to show a $\sim$50\% larger line width and size
relative to what is measured in higher-$J$ CO lines, but the values
are marginally consistent within the relative uncertainties.

The \aco\ emission in \smmjb\ shows a bright central peak
that appears to be associated with the emission that dominates the
flux in higher-$J$ CO lines. It also shows $>$2$\times$ broader
emission (as also seen in single-dish
observations at low spatial resolution; Hainline et al.\
\citeyear{hai06})  distributed over several peaks that extends out to 
$>$15\,kpc scales, but appears to be faint in higher-$J$ CO
lines. Thus, the EVLA \aco\ observations reveal a spatially resolved
CO excitation gradient in this SMG.  This may suggest that this SMG is
in an advanced stage of a merger, where the warm and dense, highly
excited gas component predominantly arises from a starburst in the
``overlap'' region of two gas-rich galaxies, and the broader, more
extended, less highly excited gas is associated with tidal structure
redistributed by the merger. More sensitive observations are required
to study the dynamical structure of this complex system in more
detail.

The \aco\ properties of \smmjb\ contrast those of high-$z$ quasars,
which appear to be dominated by less extended ($\sim$5\,kpc), highly
excited molecular gas reservoirs, with only little evidence for
low-excitation gas (e.g., Riechers et al.\ \citeyear{rie06};
\citeyear{rie09}; \citeyear{rie11}; Wei\ss\ et al.\
\citeyear{wei07}). They also contrast those of lensed $z$$\sim$3
Lyman-break galaxies, which typically have an order of magnitude less
massive, compact ($\sim$1--2\,kpc) \aco\ reservoirs (Riechers et al.\
\citeyear{rie10b}; see also higher-$J$ CO observations by Baker et
al.\ \citeyear{bak04}; Coppin et al.\ \citeyear{cop07}). Massive,
gas-rich star-forming galaxies can host gas reservoirs comparable in
extent to those found in our target SMGs, but their spatial and
velocity structure typically appears more symmetric and ordered (e.g.,
Daddi et al.\ \citeyear{dad10}). Also, their overall CO excitation
appears to be lower (Dannerbauer et al.\ \citeyear{dan09}; Aravena et
al.\ \citeyear{ara10}). Low CO excitation was also found in the
$z$=1.44 SMG HR10, which was initially selected as an Extremely Red
Object (ERO; Andreani et al.\ \citeyear{and00}; Papadopoulos \& Ivison
\citeyear{pi02}; Greve et al.\ \citeyear{gre03}).

The new \aco\ observations of the $z$$\sim$3.4 SMGs \smmja\
and \smmjb\ provide supporting evidence that they represent
early and advanced stage, gas-rich major mergers. Recent, lower
resolution \aco\ studies of the $\sim$20\,kpc separation merger
SMM\,J123707+6214 ($z$=2.488; Ivison et al.\ \citeyear{ivi11};
Riechers et al.\ \citeyear{rie11a}) may suggest an intermediate merger
stage (with two separated components, but closer in projected distance
and redshift than in \smmja ), in between those observed in
these two targets. These EVLA observations thus may be a first step
toward establishing a molecular gas-based ``merger sequence'' for
gas-rich starburst galaxies at high redshift, which may crucially
constrain formation models of SMGs (e.g., Dav\'e et al.\
\citeyear{dav10}; Hayward et al.\ \citeyear{hay11}).

If the \aco\ properties of the $z$$\sim$3.4 systems studied here are
representative of their parent population, SMGs in advanced merger
stages exhibit substantially more extended, broader low-$J$ CO line
emission than in earlier merger stages. Observations of high-$J$ CO
line emission alone will underestimate the gas content and extent of
the tidal structure in advanced-stage mergers, yielding an incomplete
picture of the processes that drive the gas dynamics in such systems.

The observations presented here thus show the key importance of
spatially and dynamically resolved \aco\ observations of SMGs to
understand the gas physics that drive star formation in these
luminous, massive gas-rich high redshift galaxies.  In combination
with higher-$J$ CO line observations with shorter wavelength
interferometers such as the upcoming Atacama Large (sub)Millimeter
Array (ALMA), this makes the EVLA a uniquely powerful tool to
distinguish different high-$z$ galaxy populations based on their
molecular gas content.

\acknowledgments 
We thank the referee, Dr.~Laura Hainline, for a helpful report, and
Christian Henkel for the original version of the LVG code. DR
acknowledges support from NASA through a Spitzer Space Telescope
grant. The National Radio Astronomy Observatory is a facility of the
National Science Foundation operated under cooperative agreement by
Associated Universities, Inc.

\end{document}